\documentclass[twocolumn,floatfix,prl,aps,showpacs]{revtex4-1}
\usepackage{graphicx,amsmath,amssymb,color}

\newcommand{\be}{\begin{equation}}
\newcommand{\ee}{\end{equation}}

\newcommand{\ba}{\begin{eqnarray}}
\newcommand{\ea}{\end{eqnarray}}

\begin{document}

\title{Fractional angular momentum in cold atom systems}
\author{Yuhe Zhang,$^1$ G. J. Sreejith,$^2$ N. D. Gemelke,$^1$ and J. K. Jain$^1$}

\affiliation{
   $^{1}$Department of Physics, 104 Davey Lab, Pennsylvania State University, University Park, PA 16802, USA}
   \affiliation{$^2$NORDITA, Roslagstullsbacken 23, 10691 Stockholm, Sweden}

\begin{abstract} The quantum statistics of bosons or fermions are manifest through even or odd relative angular momentum of a pair. We show theoretically that, under certain conditions, a pair of certain test particles immersed in a fractional quantum Hall state possesses, effectively, a fractional relative angular momentum, which can be interpreted in terms of fractional braid statistics. We propose that the fractionalization of the angular momentum can be detected directly through the measurement of the pair correlation function in rotating ultra-cold atomic systems in the fractional quantum Hall regime. Such a measurement will also provide direct evidence for the effective magnetic field resulting from Berry phases arising from attached vortices, and of excitations with fractional particle number, analogous to fractional charge of electron fractional quantum Hall effect.
\pacs{03.65.Vf,03.75.Mn,73.43.-f}
\end{abstract}
\maketitle

While all particles in nature are either bosons or fermions, emergent particles in strongly correlated condensed matter systems can, in principle, obey fractional braid statistics \cite{Leinaas77,Wilczek82}, which refers to the property that their braiding produces phases that are non-integral multiples of $2\pi$. It was proposed three decades ago \cite{Halperin84,Arovas84} that the fractional quantum Hall effect \cite{Tsui82} (FQHE) provides a platform for the realization of such entities. No convincing measurement of the fractional braid statistics has yet been made.  In this Letter we consider a pair of test atoms in a background FQHE state of bosonic atoms. Under certain conditions, the test atoms capture vortices and the bound states of atoms and vortices behave effectively as particles with fractional braid statistics. Just as fermionic or bosonic statistics are reflected through an odd or even integer relative angular momentum for a pair of particles, fractional braid statistics are manifest through fractional relative angular momentum. We further show that the relative angular momentum can be deduced from the pair correlation function through determination of the radii of various quantized orbits of one test particle around another. This provides a method for measuring fractional braid statistics relying only on already existing experimental methods of introducing test atoms as well as of measuring their pair correlation function in ultra-cold bosons in rapidly rotating optical traps.  No direct interferometric or phase measurement is necessary. In addition, our proposed experiment will provide a direct measurement of the effective magnetic field arising from a binding of vortices to bosons, as well as of excitations involving a fraction of a boson.

Neutral bosons can in principle be driven into the FQHE regime by rapid rotation, which effectively amounts to application of a magnetic field. The strongly interacting regime is reached as the number of vortices ($N_V$) in a rotating Bose-Einstein condensate becomes comparable with the number of atoms $N$, which is parametrized by the filling factor $\nu=N/N_V$.  Various methods have been developed for producing vorticity in atomic Bose  \cite{Matthews99,Madison00,Haljan01,Raman01,Lin09,Lin11} and Fermi gases~\cite{Zwierlein05}, and extremely high vorticity \cite{Ho01,Cooper08,Viefers08} (i.e. low filling factor) has been achieved~\cite{Aboshaeer01,Bretin04,Schweikhard04}. Progress toward achieving the FQHE regime has been made~\cite{Gemelke10} by implementing an adiabatic pathway to the ground state of $N$ bosons at any given total angular momentum $L$, which requires addition of a weak quadrupolar potential (of $x^2-y^2$ form) in the rotating frame to break the rotational symmetry and open gaps at the points that would otherwise be level crossings. It is possible, with standard techniques, to measure local pair correlation (by creating molecules through photoassociation), momentum distribution and the density profile of the ground state (by time of flight), local triplet correlation (by exploiting Feshbach resonance).  We propose that the new developments in high-resolution optical microscopy and single-atom detection \cite{Gemelke09,Bakr09,Sherson10} combined with a short time-of-flight expansion can be exploited to enable a direct measurement of fractional braid statistics.

Consider a situation in which two test bosons, labeled by their complex coordinates $z=x+iy$ and $z'=x'+iy'$, have been introduced into an incompressible FQHE state of $N$ bosons at $w_j=x_j+iy_j$ at filling factor $\nu$, with all particles confined to the lowest Landau level (LL).  As a concrete example, the test bosons could be Rb atoms in the internal state $F=2$, $m_F=2$, inside a FQHE droplet of Rb atoms in the internal state $F=1$, $m_F=1$. For reference, first consider the situation in which the test particles are uncorrelated with the background FQHE state of the $w$ particles. They have a wave function of the form $P_M[\{z,z'\}]e^{-|z|^2/4\ell^2-|z'|^2/4\ell^2}$, where $P_M[\{z,z'\}$ is a symmetric polynomial with angular momentum $M$ and $\ell=\sqrt{\hbar c/eB}$ is the magnetic length (related to the harmonic trap oscillator length $a_0$ by $a_0=\sqrt{2}\ell$). While many independent wave functions may be constructed for a given $M$, if we take one of the test particles at the origin ($z'=0$), which produces the most symmetric situation, all of these reduce to $P_M[\{z,z'\}]=z^M$.  
The pair wave function $z^M e^{-|z|^2/4\ell^2}$ produces a ring with a maximum density at radius $R$ given by
\begin{equation}
R^2 =2M \ell^2 \;\; {\rm (unscreened)}
\label{unsc}
\end{equation}
This can be obtained in a semiclassical approximation by demanding that the number of flux quanta ($\phi_0=hc/e$) enclosed by the circle of radius $R$, i.e. $\pi R^2 B/\phi_0=R^2/2\ell^2$, be equal to the number of enclosed vortices $M$.

We now ask how screening by the FQHE state alters this behavior. We take the combined system to be described by the following (unnormalized) wave function (suppressing here and below the ubiquitous Gaussian factor):
\begin{equation}
\Psi_{\nu}(\{ w_j\}) P_M[\{z,z'\}] \prod_{j=1}^N [(w_j-z')(w_j-z)]
\label{basis}
\end{equation}
The total angular momentum now is $L=L_{\nu}+M+2N$, where $L_{\nu}$ is the contribution of $\Psi_{\nu}(\{ w_j\})$. 
A Hamiltonian producing this wave function is given below. The crucial feature is that the repulsion between the w particles and the test particles causes a binding of unit vortices to the test particle coordinates $z$ and $z'$ through the last factor in Eq. 2.  In any adiabatic braiding process involving the test particles, the total phase accumulated is an aggregate effect of vortex and particle braiding. The Berry phase associated with braiding of vortices in a FQHE state is well-known to give rise to fractional  statistics~\cite{Arovas84}. Because bosons do not produce any additional Berry phases (modulo $2\pi$) for exchanges or windings, the bound state of a test boson and a vortex also obeys fractional braid statistics. The correlations with the $w$-particles thus produce long range gauge forces that ``screen" the statistics of the test bosons. Furthermore, by analogy to fractional Òcharge" in electronic FQHE \cite{Laughlin83}, each vortex depletes precisely $\nu$ w-atoms from its vicinity. 

We now show that the information about the fractional statistics is contained in the radii $R$ of the new rings formed by the $z$ particle (with $z'=0$), which we obtain in a mean field approximation as follows. For a given $M$, the mean number of enclosed vortices inside a loop of radius $R$ is $M+\tilde{N}=M+\nu R^2/2\ell^2-\nu$, where $\tilde{N}$ is the average number of enclosed $w$-particles, and the last term reflects the fact that the vortex tied to the centrally located test particle amounts to an expulsion of precisely $\nu$ $w$-particles from the loop. Equating it to the number of enclosed flux quanta $R^2/2 \ell^2$ gives 
\begin{eqnarray}
R^2&=& 2M_{\rm eff}  \ell^{*2}
\;\; {\rm (screened)} 
\label{fs}\\
M_{\rm eff}  &=& M-\nu \label{Meff}\\
\ell^{*2}&=&\ell^2/(1-\nu) \label{l*}
\end{eqnarray}
This result is confirmed below by explicit calculation. The two most relevant aspects are as follows. First, the size of the pair is governed by an effective magnetic length $\ell^*$ rather than $\ell$. This is a manifestation of the effective magnetic field. Because the $z$ particle sees a vortex at each $w$ particle, it experiences an effective magnetic field $B^*=B-\rho_w \phi_0=B-\nu B$ (with $\rho_w$ denoting the interior density of the $w$ particles), which, with $\ell^*\equiv \sqrt{\hbar c/e|B^*|}$, corresponds precisely to the relation given in Eq.~\ref{l*}. Second, the effective relative angular momentum $M_{\rm eff}$ is shifted relative to $M$ by an amount $\nu$ which is in general fractional. 
The fractionalization of the relative angular momentum of the test atoms is a direct manifestation of their fractional braid statistics. It is interesting that both the effective magnetic field and the fractional braid statistics can be detected directly through a measurement of the pair correlation function of the test atoms, without requiring any interferometric Berry phase measurements.

We expect Eq.~\ref{fs} to be valid so long as the $z$-particle is comfortably inside the $w$-disk. For larger $M$, when the $z$ particle lies fully outside the disk containing the $w$-particles, the number of enclosed vortices is simply $M+N$, which produces
\begin{equation}
R^2=2(M+N) \ell^2\;\;{\rm (outside)}
\label{fs2}
\end{equation}
This loses information of fractional braid statistics ($M_{\rm eff}$ has no fractional part), but retains information about correlations between the $z$ and $w$ particles through the integer shift of $M$. In deriving Eq.~\ref{fs2} we assumed that the $z$ particle is still correlated with the $w$ particles. When it is far outside the disk, the correlations disappear and the orbits revert to $R^2=2M\ell^2$ of Eq.~\ref{unsc}.

In the remainder of the article we discuss certain issues that relate to the feasibility of the measurement of the fractional braid statistics by this method.  Currently, it appears possible to obtain FQHE conditions only for relatively small systems. It is important to determine how large $N$ must be to reveal fractional braid statistics in a convincing manner, given that both test atoms must be fully inside the FQHE droplet, yet not so close that  
the overlap between the density variations associated with them begins to make substantial correction to the braid statistics parameter  \cite{Kjonsberg99,Kjonsberg99b,Jeon03b}.
To address this and certain other questions, we need wave functions for the FQHE state $\Psi_{\nu}(\{ w_j\})$ in Eq.~\ref{basis}. For $\nu=1/2$, we use Laughlin's wave function\cite{Laughlin83} $\Psi_{\nu={1\over 2}}(\{ w_j\})=\prod_{j<k}(w_j-w_k)^2$.  For general fractions of the form $\nu=n/(n\pm 1)$ we use the composite fermion (CF) wave functions \cite{Jain89,Jain07,Cooper99,Wilkin00,Regnault03,Chang05b,Korslund06,Borgh08,Wu13,Meyer14} for bosons in the LLL, given by $\Psi_{\nu={n\over n\pm 1}}(\{ w_j\})={\cal P}_{\rm LLL} \Phi_{\pm n} \prod_{j<k}(w_j-w_k)$, where $\Phi_{n}$ is the wave function of the integer quantum Hall  state with $n$ filled LLs, $\Phi_{-n}\equiv [\Phi_{n}]^*$, and ${\cal P}_{\rm LLL}$ denotes projection into the lowest LL (LLL); these represent the physics that bosons capture one vortex each to transform into composite fermions which, in turn, compactly fill $n$ CF LLs to produce incompressible FQHE states. (One filled CF LL corresponds to Laughlin's 1/2 wave function. Incidentally, the bound state of the test boson and unit vortex is also a composite fermion.)  The LLL projection will be evaluated by methods given in the literature \cite{Jain97b,Chang05b}.

\begin{figure}
\hspace{-2.1mm}
\resizebox{0.38\textwidth}{!}{\includegraphics{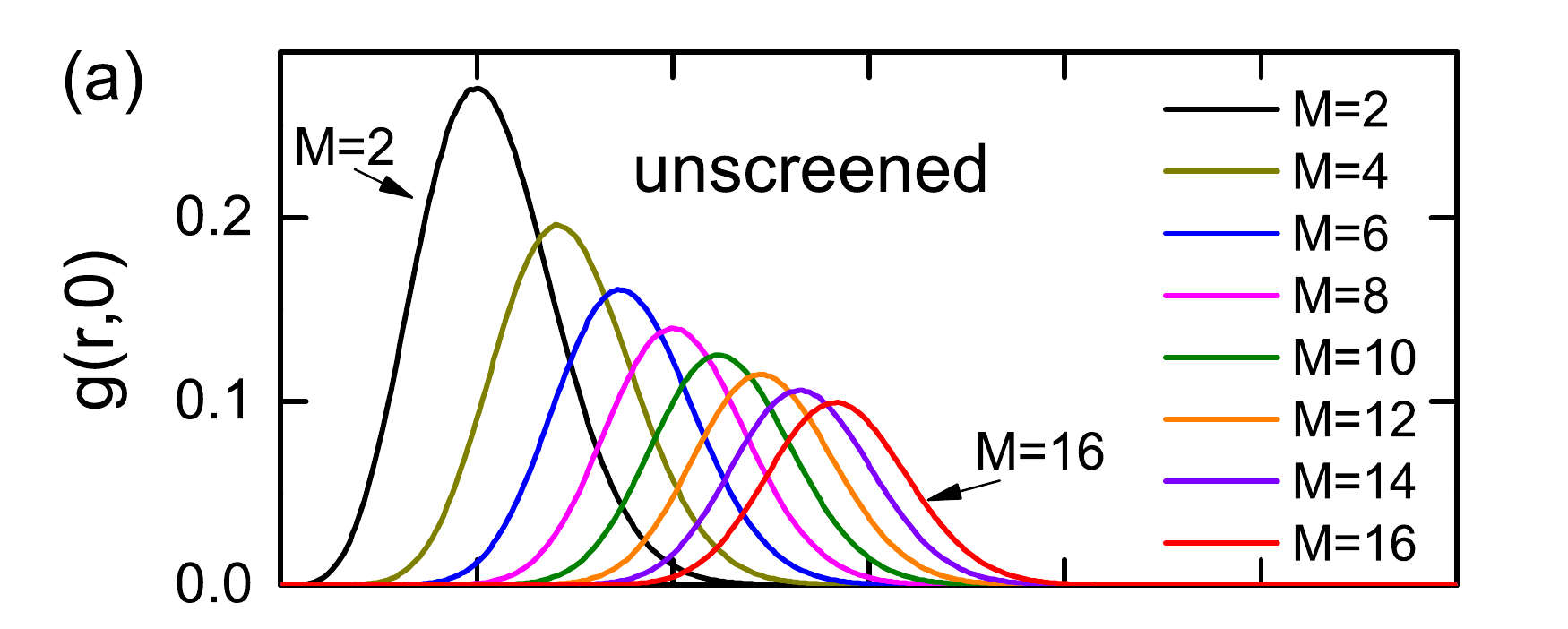}}\\
\vspace{-2.8mm}
\hspace{-3.1mm}
\resizebox{0.38\textwidth}{!}{\includegraphics{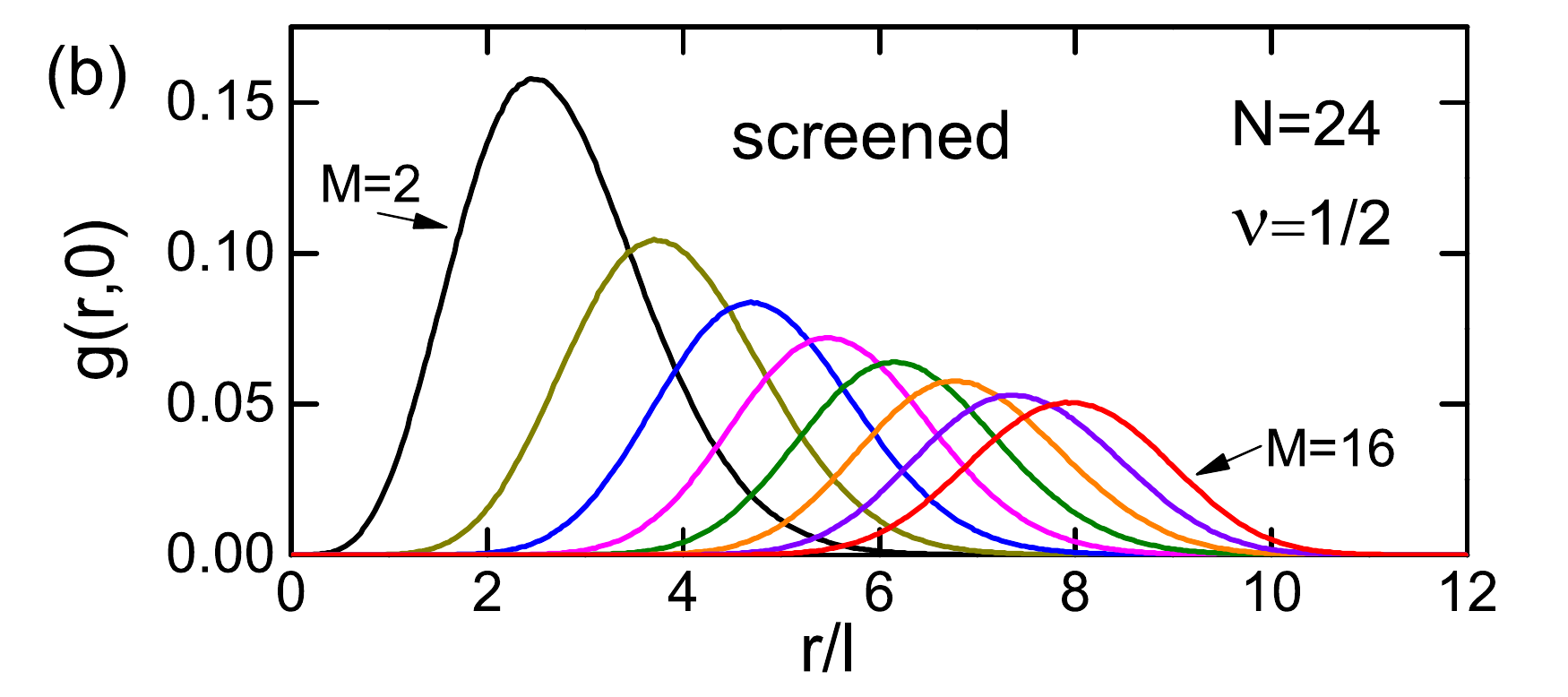}}\\
\vspace{-1mm}
\resizebox{0.245\textwidth}{!}{\includegraphics{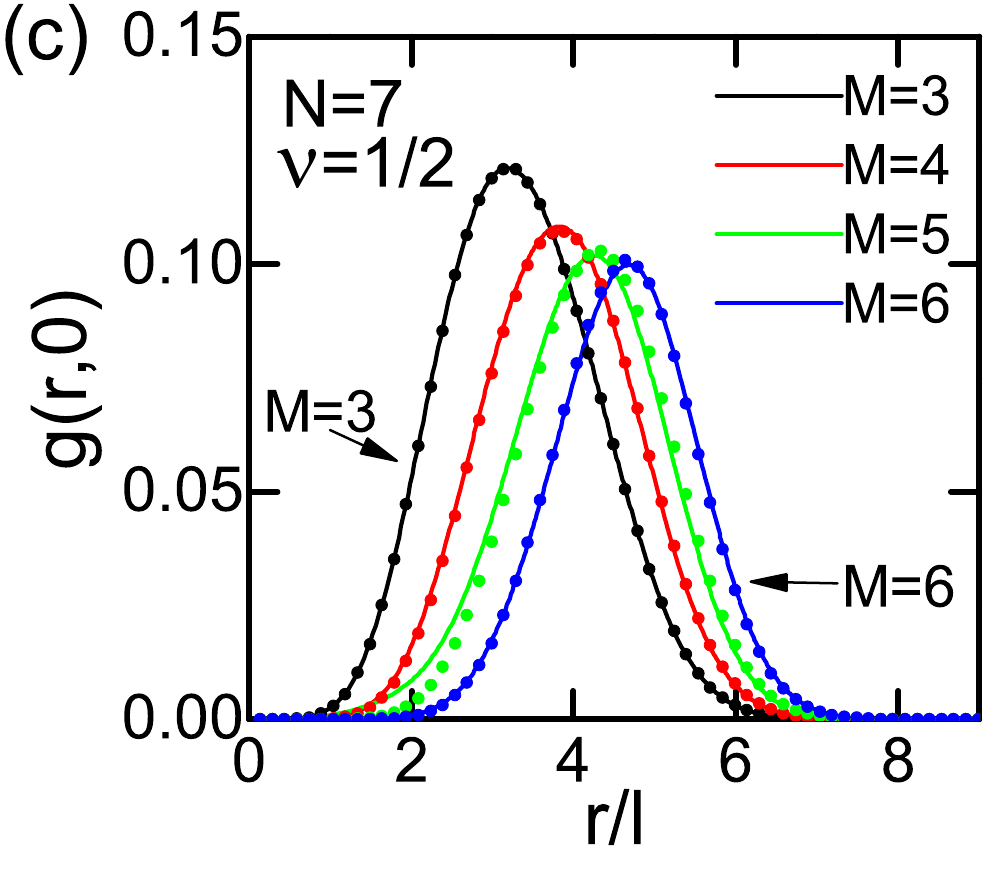}}
\hspace{-11.8mm}
\resizebox{0.245\textwidth}{!}{\includegraphics{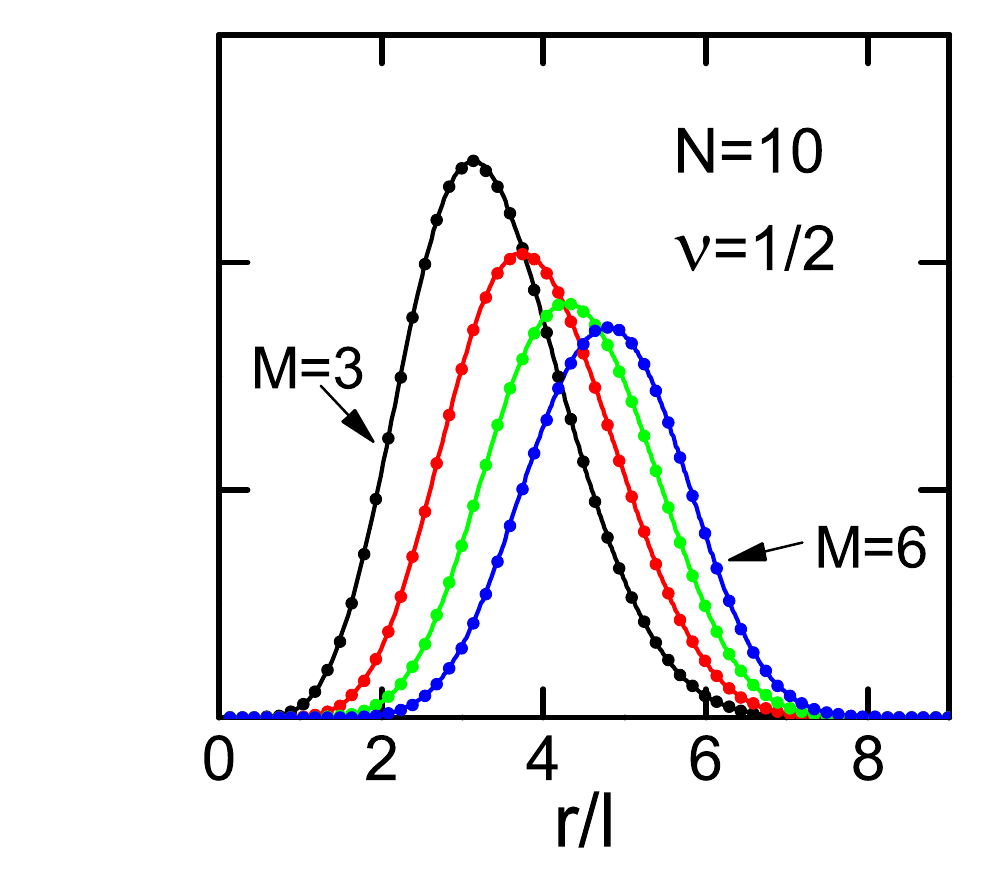}}\\
\vspace{-4mm}
\caption{(Color online). The pair correlation function $g(r,0)$ of the test particles with one particle fixed at the origin. (a) The density profile for the $z$ particle without any background $w$ particles. (b) The density profile for the $z$ particle in the background of a $\nu=1/2$ FQHE state. (c) Solid lines show the exact pair correlation function at $L=N^2+N+M$ evaluated for the the Hamiltonian in Eq.~\ref{H} assuming a high contact interaction and the condition $\omega_w>\omega_z$; dots show the pair correlation function obtained from the wave function in Eq.~\ref{basis}. The pair correlation function is quoted in units of $\rho_{0}=(2\pi {\ell}^{2})^{-1}$, where $\ell$ is the magnetic length. The number of $w$-particles, $N$, is as shown. 
\label{density}}
\end{figure}

\begin{figure}
\resizebox{0.262\textwidth}{!}{\includegraphics{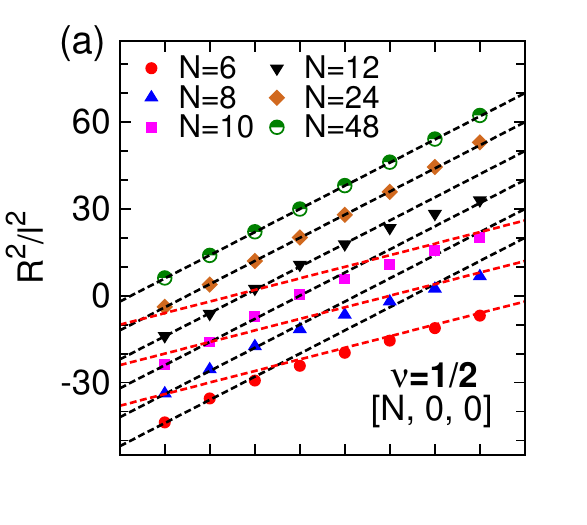}}
\hspace{-9mm}
\resizebox{0.257\textwidth}{!}{\includegraphics{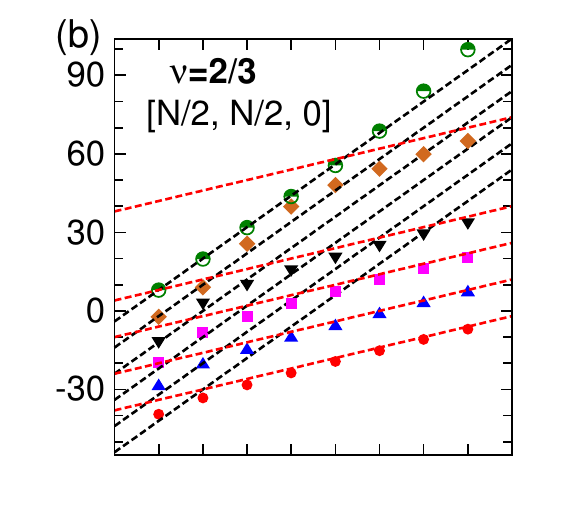}}\\
\vspace{-8.3mm}
\hspace{-1.8mm}
\resizebox{0.268\textwidth}{!}{\includegraphics{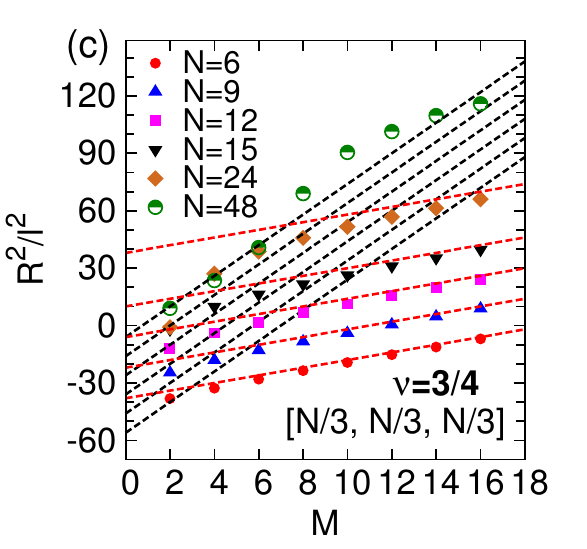}}
\hspace{-9.3mm}
\resizebox{0.259\textwidth}{!}{\includegraphics{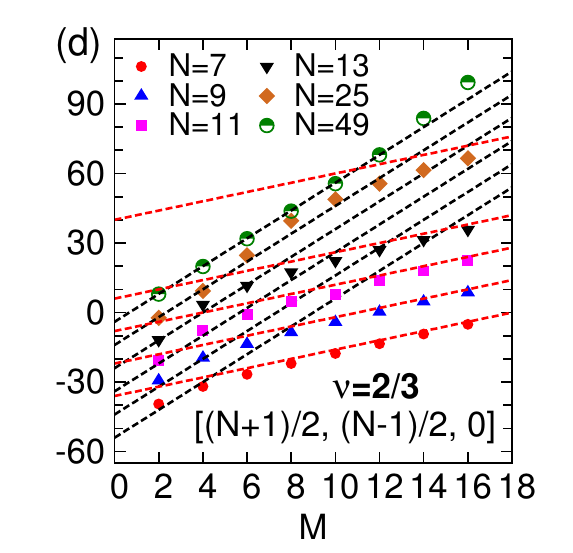}}\\
\vspace{-4mm}
\caption{(Color online). The peak position as a function of the relative angular momentum $M$ for $\nu=$1/2, 2/3 and 3/4 for several values of $N$, with $[N_0,N_1,N_2]$ representing the occupations of the lowest three CF LLs. The plot for largest $N$ is unshifted; for ease of depiction, for each successive $N$, the plot has been shifted down by 10 units. The black and red dashed lines display the behavior in Eqs.~\ref{fs} and \ref{fs2}, respectively. The fractional part of the angular momentum $M_{\rm eff}$ is given by the x-intercept of the black dashed line. The most probable separation of pairs of test atoms immersed in a fractional Hall sea of majority atoms thus demonstrates fractionalization of their relative angular momentum.} \label{Rplot}
\end{figure}

Fig.~\ref{density} (b) shows the pair correlation function $g(|z|=r,z'=0)$, i.e. the conditional density distribution of an test atom given another test atom located at the origin, as a function of $M$ for $N=24$ for $\nu=1/2$ (only even values of $M$ are shown for convenience). For comparison, the same quantity is also shown in panel (a) for a system containing only the two test particles (no $w$-particles). It is evident that the length scale governing the test pair is enhanced due to interaction with the correlated FQHE state. Fig.~\ref{Rplot} plots the peak positions as a function of $N$ and $M$ for several fractions of the form $\nu=n/(n+1)$, and compares them to the predicted behaviors in Eqs.~\ref{fs} and ~\ref{fs2}.  For 2/3 and 3/4, which map into CF filling of $\nu^*=2$ and 3, there is some ambiguity (for small systems) regarding how many composite fermions occupy each CF LL; we have assumed occupations $[N_0,N_1,N_2]$ shown in the panels (a)-(d) for the three lowest CF LLs. A number of features are notable. For sufficiently small $M$, the behavior predicted by Eq.~\ref{fs} is fully confirmed, and holds even for the smallest value $M=2$. In particular, the slope is related to the effective magnetic field, and the x-intercept gives the fractional part of the effective angular momentum. For sufficiently large $M$ the peak positions are consistent with Eq.~\ref{fs2}. The behavior changes from Eq.~\ref{fs} to Eq.~\ref{fs2} as the test atom $z$ moves, with increasing $M$, out of the disk containing the $w$-particles. Our explicit calculations show that for $\nu=1/2$ the fractionalization of angular momentum can be seen, encouragingly, even in systems with as few as 6 particles. For $\nu=2/3$ and $\nu=3/4$, larger systems are needed to reveal the fractionalization physics because, for a given $N$, the size of the FQHE droplet at $\nu=n/(n+1)$ decreases with increasing $n$. We stress that even the smallest screened pair with $M=2$ has a radius of $R=\sqrt{2n+4}\,\ell$, which  differs significantly from the radius $R=2\ell$ of an unscreened pair.

We next come to how the wave function in Eq.~\ref{basis} may be prepared for $\nu=1/2$. We consider the standard Hamiltonian in the LLL (in the rotating frame):
\be
H=V_0\sum_{j<k=1}^{N+2}\delta(\vec{r}_j-\vec{r}_k)+(\omega_w-\Omega) \hat{L}_w+(\omega_z-\Omega)  \hat{L}_z, 
\label{H}
\ee
where the contact interaction is (non-critically) assumed to be independent of the species, $\Omega$ is the rotation frequency, $\omega_w$ and $\omega_z$ are the harmonic confinement frequencies for the $w$ and $z$ particles, and $\hat{L}_w$ and $\hat{L}_z$ are total angular momentum operators for the $w$ and $z$ particles. The wave function in Eq.~\ref{basis} has zero interaction energy, but, in general, there are several zero energy states that do not have the form in Eq.~\ref{basis}. Assuming that  $(\omega_w-\Omega)/V_0\rightarrow 0$, $(\omega_z-\Omega)/V_0\rightarrow 0$, and $\omega_w\gtrsim \omega_z$, and neglecting the slight difference between the magnetic lengths of the $z$ and $w$ particles, we diagonalize the last two terms of $H$ within the basis of zero interaction energy states, and find that the exact ground state thus obtained is very well approximated by a wave function of the form given in Eq.~\ref{basis}.  This can be expected because the confinement potential preferentially suppresses edge excitations of the $w$ particles, leaving it for the test atoms fully to absorb the additional angular momentum $M$. Fig.~\ref{density}(c) shows a comparison of the pair correlation functions obtained from  the wave function in Eq.~\ref{basis} with those obtained from the exact ground state of the above Hamiltonian; the high degree of agreement confirms the validity of Eq.~\ref{basis} for our considerations. (For $M=2$ the wave function in Eq.~\ref{basis} is exact with $F_2=(z-z')^2$ in the limit $V_0\rightarrow \infty$, being the unique zero interaction energy wave function.) For general $\nu=n/(n\pm 1)$, preparation of the wave function in Eq.~\ref{basis} will require a more complicated interaction that differentiates between $w$ and $z$ particles; we leave the determination of the relevant parameter regime for a future study.

\begin{figure}
\resizebox{0.43\textwidth}{!}{\includegraphics{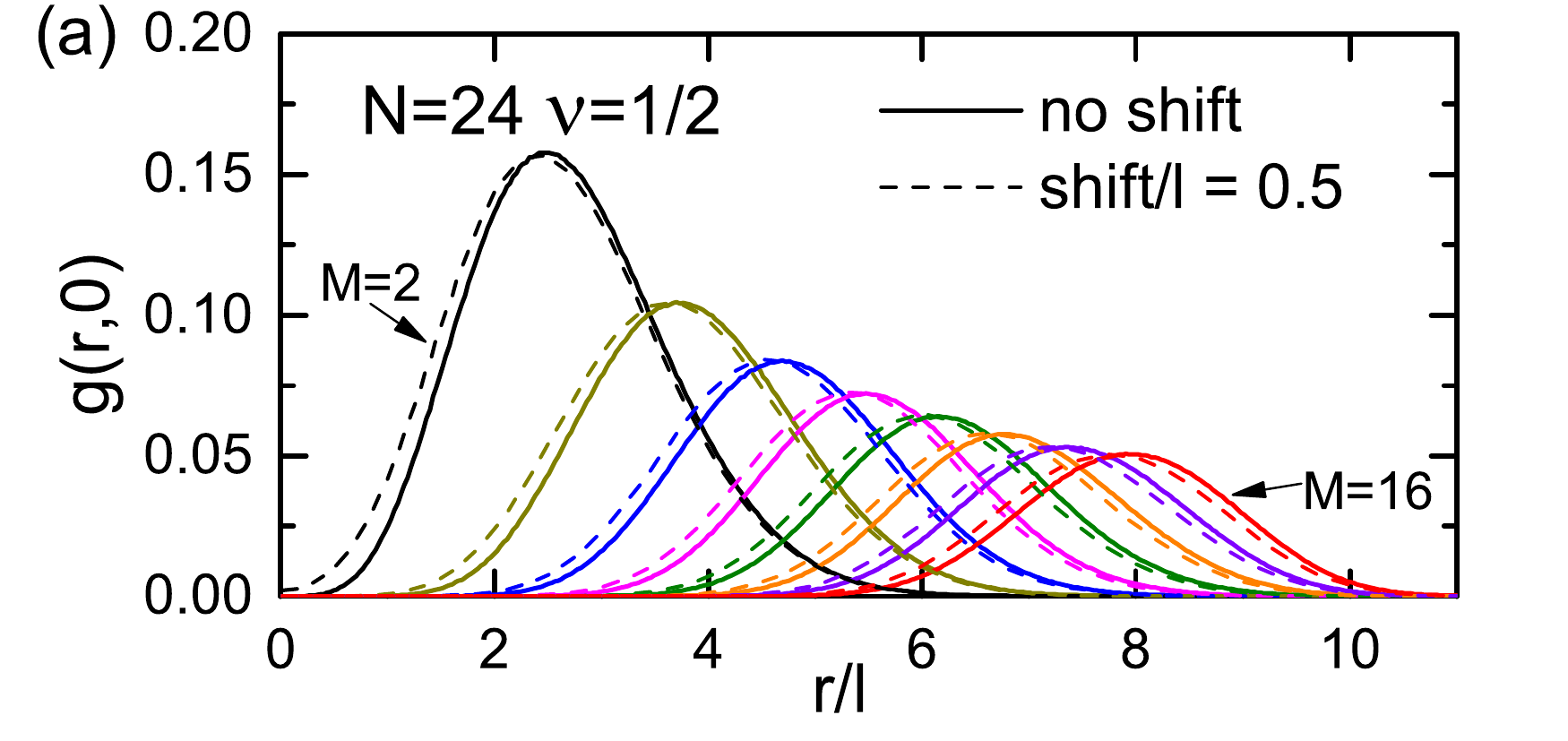}}\\
\resizebox{0.235\textwidth}{!}{\includegraphics{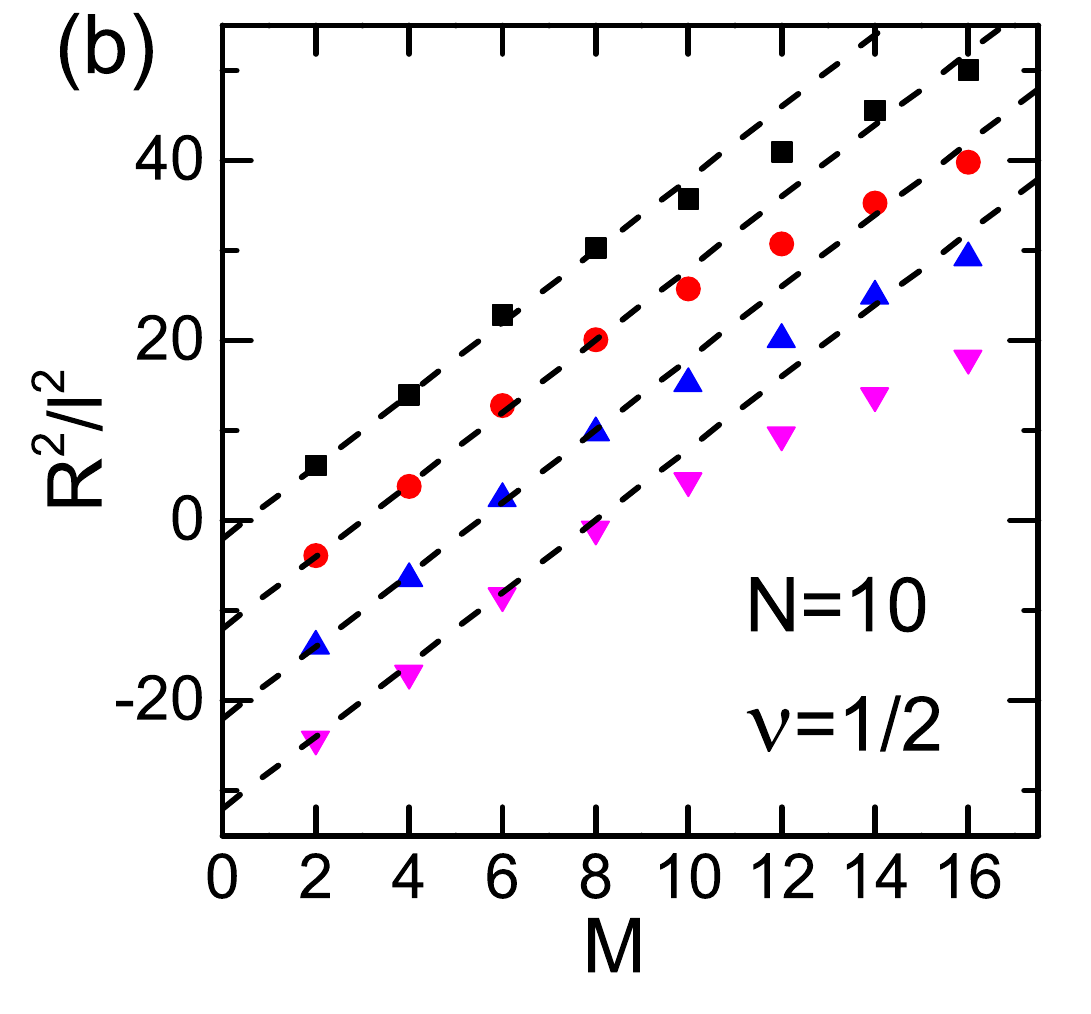}}
\hspace{-4mm}
\resizebox{0.235\textwidth}{!}{\includegraphics{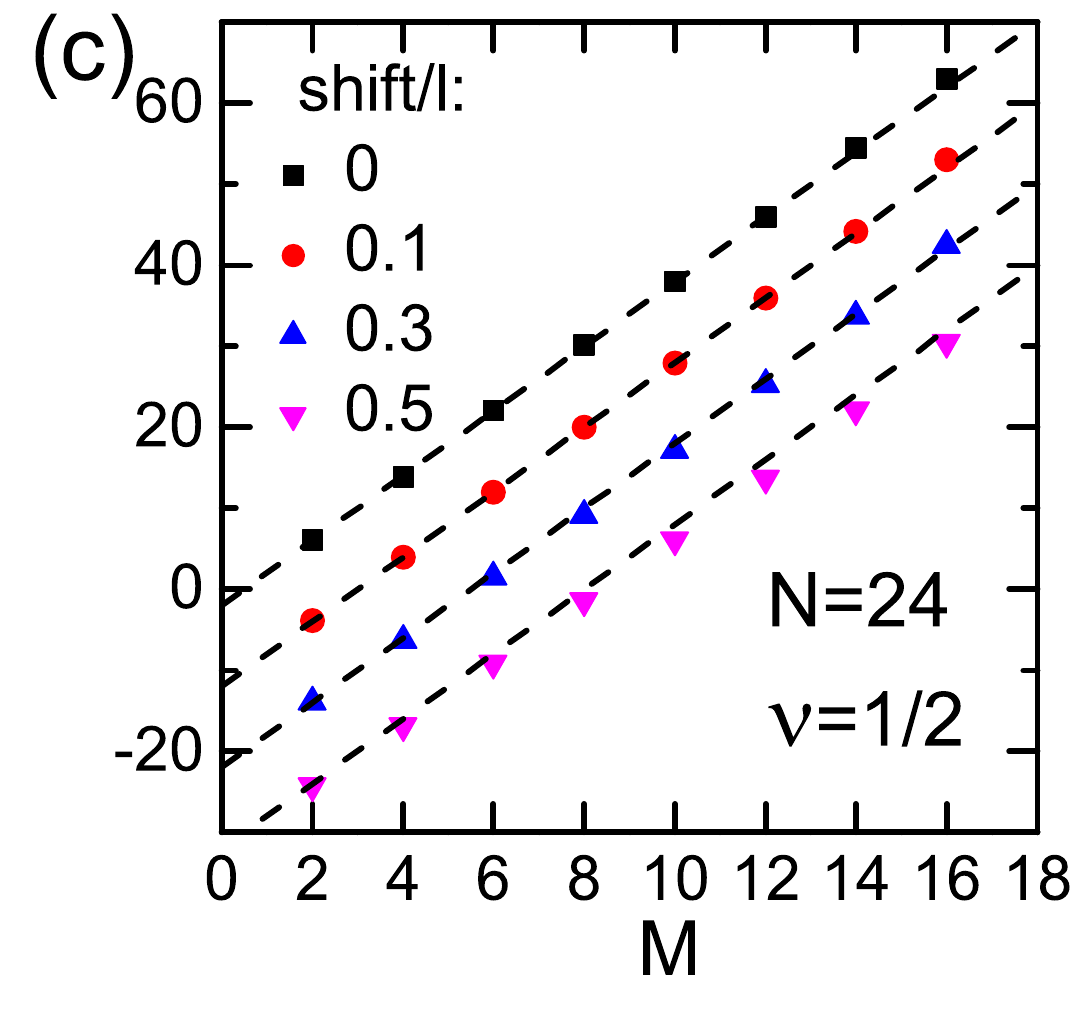}}
\vspace{-4mm}
\caption{(Color online). (a) The change in the peak position $R$ caused by shifting the test particle $z'$ off center. The colors correspond to the legend in Fig.~1a. The change is negligible for a shift of up to $0.3 \ell$. Panels (b) and (c)  show peak positions as a function of M and shift size; for clarity, the numbers for each value of shift have been shifted down by 10 units.
\label{shift}}
\end{figure}

The adiabatic scheme implemented by Gemelke {\em et al.}\cite{Gemelke10} can, in principle, be used to prepare the wave function in Eq.~\ref{basis}. This method allows one to reach the ground state at a given $L$ by an adiabatic ramp of the rotation rate provided its energy $E(L)$ satisfies $2E(L)<E(L-1)+E(L+1)$. 
Explicit evaluation of the energy $E(L)=(\omega_w-\omega_z)\langle L_w\rangle$ shows that this condition is satisfied for $M=0$ and $M=2$, and thus at a minimum, these states can be adiabatically prepared and provide useful information.  For values of $M$ that do not satisfy this condition, the ground state can in principle be prepared through extensions of the adiabatic method to prepare excited states, provided sufficient information about that spectrum is known to design and optimize the appropriate pathway.  While the methods described above are sufficient to calculate these, we defer that to future work.

It is not necessary to create highly pure states of well-defined $L$. For partially adiabatic sequences resulting in superposition states of different $L$, the distribution of conditional measurement outcomes for $r$ should reflect a weighted sum of the curves in figure \ref{density}; in order to extract data sufficient to demonstrate the fractionalization of angular momentum, it is only necessary that the expectation value $\langle L \rangle$ is known through preparation or measurement.  

In demonstrated methods for single atom detection, atoms are captured in a detection lattice which simultaneously provides confinement, cooling and detection fluorescence, and thus measurements of position are discretized to half of the detection lattice wavelength. Fractional Hall systems have previously been created using optical lattice structures angle-tuned to a larger lattice spacing, allowing for short time-of-flight expansion and reduction of density before detection. Time-of-flight expansion is self-similar in the LLL \cite{Ohberg05}, and thus measurement of the correlation functions described above is achieved in experiments by post-selecting measurement results for realizations in which one test atom is found within one detection lattice site of center. Assuming parameters similar to reference \cite{Gemelke10}, and that time-of-flight is performed sufficiently long that neighboring systems just begin to overlap, post-selection will choose systems for which $|z'|< 0.1\ell$. We have therefore also studied how the results are affected by uncertainty in the position detection by shifting the central test atom slightly off-center. As shown in Fig.~\ref{shift}, the peak positions remain essentially unchanged for shifts of up to 0.3-0.4 $\ell$ for the central test atom.

Similar experiments could detect the analog of fractional charge.  In the state described above, precisely a fractional number $\nu$ of $w$ particles is depleted from near the origin.  Direct measurement of this fractional particle number would be possible in experiments where {\em both} species of atom were imaged in-situ, and the anti-correlation of the test and majority atoms extracted over a short range. While this is perhaps difficult using existing in-situ imaging techniques (which suffer from parity detection and achieve species -- or state -- selectivity through destructive techniques), application of a strong Stern-Gerlach field gradient along the rotation axis during time-of-flight can separate the test and majority atoms into neighboring detection planes, eliminating parity issues and providing a mechanism for simultaneous detection using differential defocus of the imaging apparatus.

\underline{Acknowledgment:}  We thank Susanne Viefers for very useful discussions. We acknowledge financial support from the DOE Grant No. DE-SC0005042 (YZ, JKJ), and NSF Grant No. PHY-1068570 (NG) and thank Research Computing and Cyberinfrastructure at the Pennsylvania State University.

%

\begin{thebibliography}{38}
\expandafter\ifx\csname natexlab\endcsname\relax\def\natexlab#1{#1}\fi
\expandafter\ifx\csname bibnamefont\endcsname\relax
  \def\bibnamefont#1{#1}\fi
\expandafter\ifx\csname bibfnamefont\endcsname\relax
  \def\bibfnamefont#1{#1}\fi
\expandafter\ifx\csname citenamefont\endcsname\relax
  \def\citenamefont#1{#1}\fi
\expandafter\ifx\csname url\endcsname\relax
  \def\url#1{\texttt{#1}}\fi
\expandafter\ifx\csname urlprefix\endcsname\relax\def\urlprefix{URL }\fi
\providecommand{\bibinfo}[2]{#2}
\providecommand{\eprint}[2][]{\url{#2}}

\bibitem[{\citenamefont{Leinaas and Myrheim}(1977)}]{Leinaas77}
\bibinfo{author}{\bibfnamefont{J.}~\bibnamefont{Leinaas}} \bibnamefont{and}
  \bibinfo{author}{\bibfnamefont{J.}~\bibnamefont{Myrheim}},
  \bibinfo{journal}{Il Nuovo Cimento B Series 11}
  \textbf{\bibinfo{volume}{37}}, \bibinfo{pages}{1} (\bibinfo{year}{1977}),
  ISSN \bibinfo{issn}{0369-3554},
  \urlprefix\url{http://dx.doi.org/10.1007/BF02727953}.

\bibitem[{\citenamefont{Wilczek}(1982)}]{Wilczek82}
\bibinfo{author}{\bibfnamefont{F.}~\bibnamefont{Wilczek}},
  \bibinfo{journal}{Phys. Rev. Lett.} \textbf{\bibinfo{volume}{49}},
  \bibinfo{pages}{957} (\bibinfo{year}{1982}),
  \urlprefix\url{http://link.aps.org/doi/10.1103/PhysRevLett.49.957}.

\bibitem[{\citenamefont{Halperin}(1984)}]{Halperin84}
\bibinfo{author}{\bibfnamefont{B.~I.} \bibnamefont{Halperin}},
  \bibinfo{journal}{Phys. Rev. Lett.} \textbf{\bibinfo{volume}{52}},
  \bibinfo{pages}{1583} (\bibinfo{year}{1984}),
  \urlprefix\url{http://link.aps.org/doi/10.1103/PhysRevLett.52.1583}.

\bibitem[{\citenamefont{Arovas et~al.}(1984)\citenamefont{Arovas, Schrieffer,
  and Wilczek}}]{Arovas84}
\bibinfo{author}{\bibfnamefont{D.}~\bibnamefont{Arovas}},
  \bibinfo{author}{\bibfnamefont{J.~R.} \bibnamefont{Schrieffer}},
  \bibnamefont{and} \bibinfo{author}{\bibfnamefont{F.}~\bibnamefont{Wilczek}},
  \bibinfo{journal}{Phys. Rev. Lett.} \textbf{\bibinfo{volume}{53}},
  \bibinfo{pages}{722} (\bibinfo{year}{1984}),
  \urlprefix\url{http://link.aps.org/doi/10.1103/PhysRevLett.53.722}.

\bibitem[{\citenamefont{Tsui et~al.}(1982)\citenamefont{Tsui, Stormer, and
  Gossard}}]{Tsui82}
\bibinfo{author}{\bibfnamefont{D.~C.} \bibnamefont{Tsui}},
  \bibinfo{author}{\bibfnamefont{H.~L.} \bibnamefont{Stormer}},
  \bibnamefont{and} \bibinfo{author}{\bibfnamefont{A.~C.}
  \bibnamefont{Gossard}}, \bibinfo{journal}{Phys. Rev. Lett.}
  \textbf{\bibinfo{volume}{48}}, \bibinfo{pages}{1559} (\bibinfo{year}{1982}).

\bibitem[{\citenamefont{Matthews et~al.}(1999)\citenamefont{Matthews, Anderson,
  Haljan, Hall, Wieman, and Cornell}}]{Matthews99}
\bibinfo{author}{\bibfnamefont{M.~R.} \bibnamefont{Matthews}},
  \bibinfo{author}{\bibfnamefont{B.~P.} \bibnamefont{Anderson}},
  \bibinfo{author}{\bibfnamefont{P.~C.} \bibnamefont{Haljan}},
  \bibinfo{author}{\bibfnamefont{D.~S.} \bibnamefont{Hall}},
  \bibinfo{author}{\bibfnamefont{C.~E.} \bibnamefont{Wieman}},
  \bibnamefont{and} \bibinfo{author}{\bibfnamefont{E.~A.}
  \bibnamefont{Cornell}}, \bibinfo{journal}{Phys. Rev. Lett.}
  \textbf{\bibinfo{volume}{83}}, \bibinfo{pages}{2498} (\bibinfo{year}{1999}).

\bibitem[{\citenamefont{Madison et~al.}(2000)\citenamefont{Madison, Chevy,
  Wohlleben, and Dalibard}}]{Madison00}
\bibinfo{author}{\bibfnamefont{K.~W.} \bibnamefont{Madison}},
  \bibinfo{author}{\bibfnamefont{F.}~\bibnamefont{Chevy}},
  \bibinfo{author}{\bibfnamefont{W.}~\bibnamefont{Wohlleben}},
  \bibnamefont{and} \bibinfo{author}{\bibfnamefont{J.}~\bibnamefont{Dalibard}},
  \bibinfo{journal}{Phys. Rev. Lett.} \textbf{\bibinfo{volume}{84}},
  \bibinfo{pages}{806} (\bibinfo{year}{2000}).

\bibitem[{\citenamefont{Haljan et~al.}(2001)\citenamefont{Haljan, Coddington,
  Engels, and Cornell}}]{Haljan01}
\bibinfo{author}{\bibfnamefont{P.~C.} \bibnamefont{Haljan}},
  \bibinfo{author}{\bibfnamefont{I.}~\bibnamefont{Coddington}},
  \bibinfo{author}{\bibfnamefont{P.}~\bibnamefont{Engels}}, \bibnamefont{and}
  \bibinfo{author}{\bibfnamefont{E.~A.} \bibnamefont{Cornell}},
  \bibinfo{journal}{Phys. Rev. Lett.} \textbf{\bibinfo{volume}{87}},
  \bibinfo{pages}{210403} (\bibinfo{year}{2001}).

\bibitem[{\citenamefont{Raman et~al.}(2001)\citenamefont{Raman, Abo-Shaeer,
  Vogels, Xu, and Ketterle}}]{Raman01}
\bibinfo{author}{\bibfnamefont{C.}~\bibnamefont{Raman}},
  \bibinfo{author}{\bibfnamefont{J.~R.} \bibnamefont{Abo-Shaeer}},
  \bibinfo{author}{\bibfnamefont{J.~M.} \bibnamefont{Vogels}},
  \bibinfo{author}{\bibfnamefont{K.}~\bibnamefont{Xu}}, \bibnamefont{and}
  \bibinfo{author}{\bibfnamefont{W.}~\bibnamefont{Ketterle}},
  \bibinfo{journal}{Phys. Rev. Lett.} \textbf{\bibinfo{volume}{87}}, 
  \bibinfo{pages}{210402} (\bibinfo{year}{2001}).

\bibitem[{\citenamefont{Lin et~al.}(2009)\citenamefont{Lin, Compton,
  Jimenez-Garcia, Porto, and Spielman}}]{Lin09}
\bibinfo{author}{\bibfnamefont{Y.-J.} \bibnamefont{Lin}},
  \bibinfo{author}{\bibfnamefont{R.~L.} \bibnamefont{Compton}},
  \bibinfo{author}{\bibfnamefont{K.}~\bibnamefont{Jimenez-Garcia}},
  \bibinfo{author}{\bibfnamefont{J.~V.} \bibnamefont{Porto}}, \bibnamefont{and}
  \bibinfo{author}{\bibfnamefont{I.~B.} \bibnamefont{Spielman}},
  \bibinfo{journal}{Nature} \textbf{\bibinfo{volume}{462}},
  \bibinfo{pages}{628} (\bibinfo{year}{2009}).

\bibitem[{\citenamefont{Lin et~al.}(2011)\citenamefont{Lin, Compton,
  Jimenez-Garcia, Phillips, Porto, and Spielman}}]{Lin11}
\bibinfo{author}{\bibfnamefont{Y.~J.} \bibnamefont{Lin}},
  \bibinfo{author}{\bibfnamefont{R.~L.} \bibnamefont{Compton}},
  \bibinfo{author}{\bibfnamefont{K.}~\bibnamefont{Jimenez-Garcia}},
  \bibinfo{author}{\bibfnamefont{W.~D.} \bibnamefont{Phillips}},
  \bibinfo{author}{\bibfnamefont{J.}~\bibnamefont{Porto}}, \bibnamefont{and}
  \bibinfo{author}{\bibfnamefont{I.~B.} \bibnamefont{Spielman}},
  \bibinfo{journal}{Nature Physics} \textbf{\bibinfo{volume}{7}},
  \bibinfo{pages}{531Ð534} (\bibinfo{year}{2011}).

\bibitem[{\citenamefont{Zwierlein et~al.}(2005)\citenamefont{Zwierlein,
  Abo-Shaeer, J.~R., Schunck, and Ketterle}}]{Zwierlein05}
\bibinfo{author}{\bibfnamefont{M.~W.} \bibnamefont{Zwierlein}},
  \bibinfo{author}{\bibnamefont{Abo-Shaeer}},
  \bibinfo{author}{\bibfnamefont{A.}~\bibnamefont{J.~R.},
  \bibfnamefont{Schirotzek}}, \bibinfo{author}{\bibfnamefont{C.~H.}
  \bibnamefont{Schunck}}, \bibnamefont{and}
  \bibinfo{author}{\bibfnamefont{W.}~\bibnamefont{Ketterle}},
  \bibinfo{journal}{Nature} \textbf{\bibinfo{volume}{435}},
  \bibinfo{pages}{1047} (\bibinfo{year}{2005}).

\bibitem[{\citenamefont{Ho}(2001)}]{Ho01}
\bibinfo{author}{\bibfnamefont{T.-L.} \bibnamefont{Ho}},
  \bibinfo{journal}{Phys. Rev. Lett.} \textbf{\bibinfo{volume}{87}},
  \bibinfo{pages}{060403} (\bibinfo{year}{2001}),
  \urlprefix\url{http://link.aps.org/doi/10.1103/PhysRevLett.87.060403}.

\bibitem[{\citenamefont{Cooper}(2008)}]{Cooper08}
\bibinfo{author}{\bibfnamefont{N.~R.} \bibnamefont{Cooper}},
  \bibinfo{journal}{Advances in Physics} \textbf{\bibinfo{volume}{57}},
  \bibinfo{pages}{539} (\bibinfo{year}{2008}).

\bibitem[{\citenamefont{Viefers}(2008)}]{Viefers08}
\bibinfo{author}{\bibfnamefont{S.}~\bibnamefont{Viefers}}, \bibinfo{journal}{J.
  Phys. - Condens. Matt.} \textbf{\bibinfo{volume}{20}},
  \bibinfo{pages}{123202} (\bibinfo{year}{2008}).

\bibitem[{\citenamefont{Abo-Shaeer et~al.}(2001)\citenamefont{Abo-Shaeer,
  Raman, Vogels, and Ketterle}}]{Aboshaeer01}
\bibinfo{author}{\bibfnamefont{J.~R.} \bibnamefont{Abo-Shaeer}},
  \bibinfo{author}{\bibfnamefont{C.}~\bibnamefont{Raman}},
  \bibinfo{author}{\bibfnamefont{J.~M.} \bibnamefont{Vogels}},
  \bibnamefont{and} \bibinfo{author}{\bibfnamefont{W.}~\bibnamefont{Ketterle}},
  \bibinfo{journal}{Science} \textbf{\bibinfo{volume}{292}},
  \bibinfo{pages}{476} (\bibinfo{year}{2001}).

\bibitem[{\citenamefont{Bretin et~al.}(2004)\citenamefont{Bretin, Stock,
  Seurin, and Dalibard}}]{Bretin04}
\bibinfo{author}{\bibfnamefont{V.}~\bibnamefont{Bretin}},
  \bibinfo{author}{\bibfnamefont{S.}~\bibnamefont{Stock}},
  \bibinfo{author}{\bibfnamefont{Y.}~\bibnamefont{Seurin}}, \bibnamefont{and}
  \bibinfo{author}{\bibfnamefont{J.}~\bibnamefont{Dalibard}},
  \bibinfo{journal}{Phys. Rev. Lett.} \textbf{\bibinfo{volume}{92}}
    \bibinfo{pages}{050403} (\bibinfo{year}{2004}).

\bibitem[{\citenamefont{Schweikhard et~al.}(2004)\citenamefont{Schweikhard,
  Coddington, Engels, Mogendorff, and Cornell}}]{Schweikhard04}
\bibinfo{author}{\bibfnamefont{V.}~\bibnamefont{Schweikhard}},
  \bibinfo{author}{\bibfnamefont{I.}~\bibnamefont{Coddington}},
  \bibinfo{author}{\bibfnamefont{P.}~\bibnamefont{Engels}},
  \bibinfo{author}{\bibfnamefont{V.~P.} \bibnamefont{Mogendorff}},
  \bibnamefont{and} \bibinfo{author}{\bibfnamefont{E.~A.}
  \bibnamefont{Cornell}}, \bibinfo{journal}{Phys. Rev. Lett.}
  \textbf{\bibinfo{volume}{92}}, \bibinfo{pages}{040404}
  (\bibinfo{year}{2004}).

\bibitem[{\citenamefont{Gemelke et~al.}(2010)\citenamefont{Gemelke, Sarajlic,
  and Chu}}]{Gemelke10}
\bibinfo{author}{\bibfnamefont{N.}~\bibnamefont{Gemelke}},
  \bibinfo{author}{\bibfnamefont{E.}~\bibnamefont{Sarajlic}}, \bibnamefont{and}
  \bibinfo{author}{\bibfnamefont{S.}~\bibnamefont{Chu}} (\bibinfo{year}{2010}),
  \urlprefix\url{http://arxiv.org/abs/1007.2677}.

\bibitem[{\citenamefont{Gemelke et~al.}(2009)\citenamefont{Gemelke, Zhang,
  Hung, and Chin}}]{Gemelke09}
\bibinfo{author}{\bibfnamefont{N.}~\bibnamefont{Gemelke}},
  \bibinfo{author}{\bibfnamefont{X.}~\bibnamefont{Zhang}},
  \bibinfo{author}{\bibfnamefont{C.-L.} \bibnamefont{Hung}}, \bibnamefont{and}
  \bibinfo{author}{\bibfnamefont{C.}~\bibnamefont{Chin}},
  \bibinfo{journal}{Nature} \textbf{\bibinfo{volume}{460}},
  \bibinfo{pages}{995} (\bibinfo{year}{2009}), ISSN \bibinfo{issn}{0028-0836}.

\bibitem[{\citenamefont{Bakr et~al.}(2009)\citenamefont{Bakr, Gillen, Peng,
  Foelling, and Greiner}}]{Bakr09}
\bibinfo{author}{\bibfnamefont{W.~S.} \bibnamefont{Bakr}},
  \bibinfo{author}{\bibfnamefont{J.~I.} \bibnamefont{Gillen}},
  \bibinfo{author}{\bibfnamefont{A.}~\bibnamefont{Peng}},
  \bibinfo{author}{\bibfnamefont{S.}~\bibnamefont{Foelling}}, \bibnamefont{and}
  \bibinfo{author}{\bibfnamefont{M.}~\bibnamefont{Greiner}},
  \bibinfo{journal}{Nature} \textbf{\bibinfo{volume}{462}}, \bibinfo{pages}{74}
  (\bibinfo{year}{2009}), ISSN \bibinfo{issn}{0028-0836}.

\bibitem[{\citenamefont{Sherson et~al.}(2010)\citenamefont{Sherson, Weitenberg,
  Endres, Cheneau, Bloch, and Kuhr}}]{Sherson10}
\bibinfo{author}{\bibfnamefont{J.~F.} \bibnamefont{Sherson}},
  \bibinfo{author}{\bibfnamefont{C.}~\bibnamefont{Weitenberg}},
  \bibinfo{author}{\bibfnamefont{M.}~\bibnamefont{Endres}},
  \bibinfo{author}{\bibfnamefont{M.}~\bibnamefont{Cheneau}},
  \bibinfo{author}{\bibfnamefont{I.}~\bibnamefont{Bloch}}, \bibnamefont{and}
  \bibinfo{author}{\bibfnamefont{S.}~\bibnamefont{Kuhr}},
  \bibinfo{journal}{Nature} \textbf{\bibinfo{volume}{467}}, \bibinfo{pages}{68}
  (\bibinfo{year}{2010}), ISSN \bibinfo{issn}{0028-0836}.

\bibitem[{\citenamefont{Laughlin}(1983)}]{Laughlin83}
\bibinfo{author}{\bibfnamefont{R.~B.} \bibnamefont{Laughlin}},
  \bibinfo{journal}{Phys. Rev. Lett.} \textbf{\bibinfo{volume}{50}},
  \bibinfo{pages}{1395} (\bibinfo{year}{1983}).

\bibitem[{\citenamefont{Jeon et~al.}(2003)\citenamefont{Jeon, Graham, and
  Jain}}]{Jeon03b}
\bibinfo{author}{\bibfnamefont{G.~S.} \bibnamefont{Jeon}},
  \bibinfo{author}{\bibfnamefont{K.~L.} \bibnamefont{Graham}},
  \bibnamefont{and} \bibinfo{author}{\bibfnamefont{J.~K.} \bibnamefont{Jain}},
  \bibinfo{journal}{Phys. Rev. Lett.} \textbf{\bibinfo{volume}{91}},
  \bibinfo{pages}{036801} (\bibinfo{year}{2003}),
  \urlprefix\url{http://link.aps.org/doi/10.1103/PhysRevLett.91.036801}.

\bibitem[{\citenamefont{Kj{\o}nsberg and Myrheim}(1999)}]{Kjonsberg99}
\bibinfo{author}{\bibfnamefont{H.}~\bibnamefont{Kj{\o}nsberg}}
  \bibnamefont{and} \bibinfo{author}{\bibfnamefont{J.}~\bibnamefont{Myrheim}},
  \bibinfo{journal}{International Journal of Modern Physics A}
  \textbf{\bibinfo{volume}{14}}, \bibinfo{pages}{537} (\bibinfo{year}{1999}).

\bibitem[{\citenamefont{Kj{\o}nsberg and Leinaas}(1999)}]{Kjonsberg99b}
\bibinfo{author}{\bibfnamefont{H.}~\bibnamefont{Kj{\o}nsberg}}
  \bibnamefont{and} \bibinfo{author}{\bibfnamefont{J.~M.}
  \bibnamefont{Leinaas}}, \bibinfo{journal}{Nucl. Phys. B}
  \textbf{\bibinfo{volume}{559}}, \bibinfo{pages}{705} (\bibinfo{year}{1999}).

\bibitem[{\citenamefont{Jain}(2007)}]{Jain07}
\bibinfo{author}{\bibfnamefont{J.~K.} \bibnamefont{Jain}},
  \emph{\bibinfo{title}{Composite Fermions}} (\bibinfo{publisher}{Cambridge
  University Press, New York, US}, \bibinfo{year}{2007}).

\bibitem[{\citenamefont{Jain}(1989)}]{Jain89}
\bibinfo{author}{\bibfnamefont{J.~K.} \bibnamefont{Jain}},
  \bibinfo{journal}{Phys. Rev. Lett.} \textbf{\bibinfo{volume}{63}},
  \bibinfo{pages}{199} (\bibinfo{year}{1989}).

\bibitem[{\citenamefont{Cooper and Wilkin}(1999)}]{Cooper99}
\bibinfo{author}{\bibfnamefont{N.~R.} \bibnamefont{Cooper}} \bibnamefont{and}
  \bibinfo{author}{\bibfnamefont{N.~K.} \bibnamefont{Wilkin}},
  \bibinfo{journal}{Phys. Rev. B} \textbf{\bibinfo{volume}{60}},
  \bibinfo{pages}{R16279} (\bibinfo{year}{1999}),
  \urlprefix\url{http://link.aps.org/doi/10.1103/PhysRevB.60.R16279}.

\bibitem[{\citenamefont{Wilkin and Gunn}(2000)}]{Wilkin00}
\bibinfo{author}{\bibfnamefont{N.~K.} \bibnamefont{Wilkin}} \bibnamefont{and}
  \bibinfo{author}{\bibfnamefont{J.~M.~F.} \bibnamefont{Gunn}},
  \bibinfo{journal}{Phys. Rev. Lett.} \textbf{\bibinfo{volume}{84}},
  \bibinfo{pages}{6} (\bibinfo{year}{2000}),
  \urlprefix\url{http://link.aps.org/doi/10.1103/PhysRevLett.84.6}.

\bibitem[{\citenamefont{Regnault and Jolicoeur}(2003)}]{Regnault03}
\bibinfo{author}{\bibfnamefont{N.}~\bibnamefont{Regnault}} \bibnamefont{and}
  \bibinfo{author}{\bibfnamefont{T.}~\bibnamefont{Jolicoeur}},
  \bibinfo{journal}{Phys. Rev. Lett.} \textbf{\bibinfo{volume}{91}},
  \bibinfo{pages}{030402} (\bibinfo{year}{2003}),
  \urlprefix\url{http://link.aps.org/doi/10.1103/PhysRevLett.91.030402}.

\bibitem[{\citenamefont{Chang et~al.}(2005)\citenamefont{Chang, Regnault,
  Jolicoeur, and Jain}}]{Chang05b}
\bibinfo{author}{\bibfnamefont{C.-C.} \bibnamefont{Chang}},
  \bibinfo{author}{\bibfnamefont{N.}~\bibnamefont{Regnault}},
  \bibinfo{author}{\bibfnamefont{T.}~\bibnamefont{Jolicoeur}},
  \bibnamefont{and} \bibinfo{author}{\bibfnamefont{J.~K.} \bibnamefont{Jain}},
  \bibinfo{journal}{Phys. Rev. A} \textbf{\bibinfo{volume}{72}},
  \bibinfo{pages}{013611} (\bibinfo{year}{2005}),
  \urlprefix\url{http://link.aps.org/doi/10.1103/PhysRevA.72.013611}.

\bibitem[{\citenamefont{Korslund and Viefers}(2006)}]{Korslund06}
\bibinfo{author}{\bibfnamefont{M.~N.} \bibnamefont{Korslund}} \bibnamefont{and}
  \bibinfo{author}{\bibfnamefont{S.}~\bibnamefont{Viefers}},
  \bibinfo{journal}{Phys. Rev. A} \textbf{\bibinfo{volume}{73}},
  \bibinfo{pages}{063602} (\bibinfo{year}{2006}),
  \urlprefix\url{http://link.aps.org/doi/10.1103/PhysRevA.73.063602}.

\bibitem[{\citenamefont{Borgh et~al.}(2008)\citenamefont{Borgh, Koskinen,
  Christensson, Manninen, and Reimann}}]{Borgh08}
\bibinfo{author}{\bibfnamefont{M.}~\bibnamefont{Borgh}},
  \bibinfo{author}{\bibfnamefont{M.}~\bibnamefont{Koskinen}},
  \bibinfo{author}{\bibfnamefont{J.}~\bibnamefont{Christensson}},
  \bibinfo{author}{\bibfnamefont{M.}~\bibnamefont{Manninen}}, \bibnamefont{and}
  \bibinfo{author}{\bibfnamefont{S.~M.} \bibnamefont{Reimann}},
  \bibinfo{journal}{Phys. Rev. A} \textbf{\bibinfo{volume}{77}},
  \bibinfo{pages}{033615} (\bibinfo{year}{2008}),
  \urlprefix\url{http://link.aps.org/doi/10.1103/PhysRevA.77.033615}.

\bibitem[{\citenamefont{Wu and Jain}(2013)}]{Wu13}
\bibinfo{author}{\bibfnamefont{Y.-H.} \bibnamefont{Wu}} \bibnamefont{and}
  \bibinfo{author}{\bibfnamefont{J.~K.} \bibnamefont{Jain}},
  \bibinfo{journal}{Phys. Rev. B} \textbf{\bibinfo{volume}{87}},
  \bibinfo{pages}{245123} (\bibinfo{year}{2013}),
  \urlprefix\url{http://link.aps.org/doi/10.1103/PhysRevB.87.245123}.

\bibitem[{\citenamefont{Meyer et~al.}(2014)\citenamefont{Meyer, Sreejith, and
  Viefers}}]{Meyer14}
\bibinfo{author}{\bibfnamefont{M.~L.} \bibnamefont{Meyer}},
  \bibinfo{author}{\bibfnamefont{G.~J.} \bibnamefont{Sreejith}},
  \bibnamefont{and} \bibinfo{author}{\bibfnamefont{S.}~\bibnamefont{Viefers}},
  \bibinfo{journal}{Phys. Rev. A} \textbf{\bibinfo{volume}{89}},
  \bibinfo{pages}{043625} (\bibinfo{year}{2014}),
  \urlprefix\url{http://link.aps.org/doi/10.1103/PhysRevA.89.043625}.

\bibitem[{\citenamefont{Jain and Kamilla}(1997)}]{Jain97b}
\bibinfo{author}{\bibfnamefont{J.~K.} \bibnamefont{Jain}} \bibnamefont{and}
  \bibinfo{author}{\bibfnamefont{R.~K.} \bibnamefont{Kamilla}},
  \bibinfo{journal}{Phys. Rev. B} \textbf{\bibinfo{volume}{55}},
  \bibinfo{pages}{R4895} (\bibinfo{year}{1997}).

\bibitem[{\citenamefont{\"Ohberg et~al.}(2005)\citenamefont{\"Ohberg,
  Juzeliunas, Ruseckas, and Fleischhauer}}]{Ohberg05}
\bibinfo{author}{\bibfnamefont{P.}~\bibnamefont{\"Ohberg}},
  \bibinfo{author}{\bibfnamefont{G.}~\bibnamefont{Juzeliunas}},
  \bibinfo{author}{\bibfnamefont{J.}~\bibnamefont{Ruseckas}}, \bibnamefont{and}
  \bibinfo{author}{\bibfnamefont{M.}~\bibnamefont{Fleischhauer}},
  \bibinfo{journal}{Phys. Rev. A} \textbf{\bibinfo{volume}{72}},
  \bibinfo{pages}{053632} (\bibinfo{year}{2005}),
  \urlprefix\url{http://link.aps.org/doi/10.1103/PhysRevA.72.053632}.

\end{thebibliography}
%
%
%
%
%

\end{document}